\documentclass[12pt,a4paper]{article}
\usepackage{amsfonts,latexsym}
\usepackage{graphicx,color}
\usepackage{dcolumn}
\usepackage{graphicx}
\usepackage{amsmath}
\usepackage{amsfonts}
\usepackage{amssymb}

\renewcommand{\thefootnote}{\fnsymbol{footnote}}

\begin{document}

\vspace{12mm}

\begin{center}
{{{\Large {\bf Scalarized black holes in new massive gravity dressed by a nonminimally coupled scalar  }}}}\\[10mm]

{Yun Soo Myung$^a$\footnote{e-mail address: ysmyung@inje.ac.kr}}\\[8mm]

{${}^a$Institute of Basic Sciences and Department  of Computer Simulation, Inje University, \\ Gimhae 50834, Korea\\[0pt] }

\end{center}
\vspace{2mm}

\begin{abstract}
We investigate  scalarized black holes in new massive gravity dressed by a nonminimally coupled scalar.
For this purpose, we find the Gregory-Laflamme (GL) and tachyonic instability bounds of bald BTZ black hole expressed in terms of $ m^2$ a massive spin-2 parameter and  $\alpha$ a scalar coupling parameter to Ricci scalar by making use of the linearized theory around black hole.
On the other hand, we obtain a solution bound of $0.161<\alpha <3/16$ for achieving non-BTZ black holes with scalar hair analytically  and a thermodynamic bound of $1/8<\alpha <13/80$ for obtaining consistent thermodynamic quantities.
Without imposing the GL instability bound ($m^2<1/2\ell^2$),  we find a very narrow  bound of $0.161<\alpha<0.1625$ which is located inside tachyonic instability bound of $\alpha>1/8+m^2_\Phi \ell^2/6$ for obtaining scalarized black holes where $m^2_\Phi$ is a scalar mass parameter.

\end{abstract}
\vspace{5mm}

\newpage
\renewcommand{\thefootnote}{\arabic{footnote}}
\setcounter{footnote}{0}

%%%% Introduction %%%%

\section{Introduction}

Recently, there was a significant progress on obtaining black holes with scalar hair called as spontaneous scalarization~\cite{Doneva:2022ewd}. This corresponds to an explicit example for evading  no-hair theorem.
In this direction,  the tachyonic instability of bald black holes is regarded as  the onset of  scalarized black holes when introducing a scalar coupling $f(\phi)$ to the source term: the Gauss-Bonnet (GB) term for Schwarzschild black hole with mass $M$~\cite{Doneva:2017bvd,Silva:2017uqg,Antoniou:2017acq} and  Kerr black hole~\cite{Cunha:2019dwb} or Maxwell term  for Reissner-Norstr\"{o}m (RN) black hole with mass mass $M$ and charge $Q$~\cite{Herdeiro:2018wub} and  Kerr-Newman  black hole~\cite{Lai:2022ppn}. For the GB coupling, the coupling function was usually chosen to be either $f(\phi)=(1-e^{-6\phi^2})/6$ or $\phi^2$ which leads to the same form when linearizing equations.   Explicitly, the tachyonic instability for a scalar perturbation can be  found  when the negative region of  potential is developed in the near horizon by including a negative mass term $\mu^2_{\rm S}=-2\alpha\bar{
R}^2_{\rm GB}=-\frac{96\alpha M^2
}{r^6}$  with a positive scalar coupling parameter $\alpha$  in the Einstein-Gauss-Bonnet-scalar (EGBS) theory or $\mu^2_{\rm RN}=\alpha \bar{F}^2/2=-\frac{\alpha (Mq)^2}{r^4}$ with $q=Q/M$ in Einstein-Maxwell-scalar theory. Another source terms might be  geometric invariant sources of Ricci scalar ($R$) and Chern-Simons term (${}^*RR$). The first branch of scalarized black holes crosses the Schwarzschild black hole at the threshold ($\alpha_{th}=0.363$ for $M=1$) of tachyonic instability, while  the first branch of scalarized charged black holes crosses the RN black hole at the threshold ($\alpha_{th}=8.45$ for $q=0.7$) of tachyonic instability.
 However, it is worth noting that  most of scalarized black hole solutions were constructed numerically because their analytic solutions are hardly found. This implies that a completely thermodynamic  study of these scalarized black holes is handicapped. For this purpose, it is desirable  to find an analytic black hole solution with scalar hair.

On the other hand,  a fourth-order gravity [Einstein-Weyl theory: $R-\frac{1}{2m_2^2}(R_{\mu\nu}^2-R^2/3)$] with a massive spin-2 parameter $m^2_2$ has  provided  the non-Schwarzschild  solution which crosses the Schwarzschild  solution at the threshold  point ($m_2^{th}=0.876$ for $r_+=2M=1$)~\cite{Lu:2015cqa}.
This numerical solution represents a black hole with  Ricci tensor hair, comparing to  Schwarzschild  with zero Ricci tensor.
Even though its approximate analytic solutions were found by making use of the continued-fraction expansion~\cite{Kokkotas:2017zwt}, one could not find any analytic form of  non-Schwarzschild solution.
At this stage, we note that the instability of Schwarzschild black hole was found in the dRGT massive gravity~\cite{Babichev:2013una,Brito:2013wya} and the instability bound of Schwarzschild black hole was  found as $m_2<0.876/r_+$  when solving  the Lichnerowicz equation for the linearized Ricci tensor $\delta R_{\mu\nu}$~\cite{Myung:2013doa}. It corresponds exactly to the linearized Einstein equation for $h_{\mu\nu}^{(4)}$ a four-dimensional metric tensor  around a five-dimensional black string where the Gregory-Laflamme (GL) instability was firstly observed~\cite{Gregory:1993vy}. More recently, it was shown that  the long-wave length  instability bound for non-Schwarzschild solution is given by $m_2<0.876/r_+$~\cite{Held:2022abx},  which is the same bound as  the GL instability for Schwarzschild solution.  However, this instability bound is not consistent with that predicted by thermodynamic analysis of non-Schwarzschild solution~\cite{Lu:2017kzi}.

Now, it is very interesting to  note that the new  massive gravity with a positive massive spin-2 parameter $m^2$~\cite{Bergshoeff:2009hq} is considered as a three-dimensional version of Einstin-Weyl gravity with a negative cosmological constant.
In this case, the GL instability bound was found to be $m^2<\frac{1}{2\ell^2}$ with $\ell$ the AdS$_3$ curvature radius~\cite{Moon:2013lea}. Also, it indicates a clear connection (CSC: correlated stability conjecture) between thermodynamic instability and
GL instability for the BTZ black hole regardless of the horizon radius $r_+$~\cite{Myung:2013uka}.  On later, analytic black hole solutions were found in  new massive gravity dressed by a nonminimally coupled scalar to Ricci scalar $R$~\cite{Correa:2014ika}. However, an interpretation of asymptotically AdS black hole solutions is not still clear  as well as  their thermodynamic analysis is incomplete. Furthermore, these solutions have received  less attention than asymptotically Lifshitz black hole solutions in three dimensions~\cite{Ayon-Beato:2015jga,Bravo-Gaete:2020ftn}.

In the present work, we  wish to revisit these AdS black hole  solutions as non-BTZ black holes with scalar hair and find all thermodynamic quantities.
Without imposing the GL instability bound,  we find a very narrow  bound of $0.161<\alpha<0.1625$ which is located inside tachyonic instability bound of $\alpha>\alpha_{th}$ with $\alpha_{th}=1/8+m^2_\Phi \ell^2/6$ for obtaining scalarized black holes by analyzing both  solution and thermodynamic bounds. On the other hand, imposing the GL instability bound, one could not find any consistent bound.

\section{BTZ black holes}

We start with the new massive gravity dressed by a nonmimally coupled scalar (NMGdbs)~\cite{Correa:2014ika}
\begin{eqnarray}
\label{NMGAct}
 S_{\rm NMGdbs} &=&S_{\rm NMG}+S_{\Phi },  \\
 \label{NEH} S_{\rm NMG}&=& \frac{1}{2} \int d^3x \sqrt{-g}~
  \Big[R-2\Lambda-\frac{1}{m^2}\Big(R_{\rho\sigma}R^{\rho\sigma}-\frac{3}{8}R^2\Big)\Big], \\
\label{NFO} S_{\Phi }&=&-\frac{1}{2} \int d^3x
            \sqrt{-g}~\Big[(\partial\Phi)^2+\alpha R \Phi^2+m_{\Phi}^2\Phi^2+\frac{\lambda}{12}\Phi^4 \Big],
\end{eqnarray}
where $m^2$ is a
positive massive spin-2 parameter,  $\alpha$ is a positive scalar coupling to Ricci scalar $R$, and $m^2_\Phi$ is a scalar mass parameter [$m,m_\Phi, \alpha \in
(0,\infty)$]. Here, it is important to note that  we introduce a quadratic scalar coupling $\phi^2$ to Ricci scalar  $R$ but not to fourth-order term $K=R_{\rho\sigma}R^{\rho\sigma}-3R^2/8$. The reason for this choice  is that the former is easy to find an analytic black hole solution  with scalar hair  than the latter.

 The Einstein equation is given by \begin{equation} \label{eqn}
G_{\mu\nu}+\Lambda g_{\mu\nu}-\frac{1}{2m^2}K_{\mu\nu}-T_{\mu\nu}^\Phi=0,
\end{equation}
where
\begin{eqnarray}
  K_{\mu\nu}&=&2\square R_{\mu\nu}-\frac{1}{2}\nabla_\mu \nabla_\nu R-\frac{1}{2}\square Rg_{\mu\nu}  \nonumber \\
        &+&4R_{\mu\rho\nu\sigma}R^{\rho\sigma} -\frac{3}{2}RR_{\mu\nu}-R_{\rho\sigma}R^{\rho\sigma}g_{\mu\nu}
         +\frac{3}{8}R^2g_{\mu\nu} \label{kmunu}
\end{eqnarray}
and the stress tensor for a scalar $\Phi$ is defined by
\begin{eqnarray}
T_{\mu\nu}^\Phi&=& \partial_\mu \Phi \partial_\nu \Phi -\frac{g_{\mu\nu}}{2}\Big[(\partial\Phi)^2+m_{\Phi}^2\Phi^2+\frac{\lambda}{12}\Phi^4 \Big]\nonumber \\
&+& \alpha\Big(g_{\mu\nu}\square-\nabla_\mu \nabla_\nu +G_{\mu\nu}\Big) \Phi^2. \label{tmunu}
\end{eqnarray}
On the other hand, the scalar equation takes the form as
\begin{equation}
\square \Phi-\alpha R\Phi -m^2_{\Phi}\Phi-\frac{\lambda}{6}\Phi^3=0. \label{sca-eq}
\end{equation}
Our primary concern is the non-rotating BTZ  black hole solution to Eqs.(\ref{eqn}) and (\ref{sca-eq}) given
by
 \begin{eqnarray}
\label{btz-m}
  ds^2_{\rm BTZ}&=&\bar{g}_{\mu\nu}dx^\mu dx^\nu=-f(r)dt^2
   +\frac{dr^2}{f(r)}+r^2d\varphi^2,~~f(r)=-M+\frac{r^2}{\ell^2}, \\
   \Phi&=&0 \nonumber
\end{eqnarray}
under the condition of $1/\ell^2+\Lambda+1/(4m^2\ell^4)=0$.
 Here $M$
is an integration constant related to the the ADM mass of BTZ black
hole  and $\ell$ denotes the curvature radius of AdS$_3$
spacetimes. The horizon radius $r_+=\sqrt{M}\ell$ is determined by the condition of
$f(r)=0$.
Its Hawking  temperature is found to be  \begin{equation} T_{\rm H}(r_+)=
\frac{f'(r_+)}{4\pi}=\frac{r_+}{2\pi \ell^2}. \end{equation} Using the Abbott-Deser-Tekin
method~\cite{Abbott:1981ff,Deser:2002rt}, one can derive  all ($m^2$-dependent) thermodynamic quantities of  its
mass, heat capacity ($C=\frac{dM_{\rm
ADT}}{dT_{\rm H}}$), entropy, and
Helmholtz free energy as~\cite{Myung:2013uka}
\begin{eqnarray} \label{lbh2} M_{\rm
ADT}=\frac{{\cal M}^2}{m^2}M(r_+),~~C_{\rm ADT}=\frac{{\cal
M}^2}{m^2}C(r_+),~~ S_{\rm ADT}=\frac{{\cal M}^2}{m^2}S_{\rm
BH}(r_+),~~F_{\rm ADT}=\frac{{\cal M}^2}{m^2}F(r_+),\end{eqnarray}
where ${\cal M}^2$ is defined by
\begin{equation}
{\cal M}^2(m^2)=m^2-\frac{1}{2\ell^2}. \label{spin2-m}
 \end{equation}
Here, thermodynamic quantities in Einstein gravity are
given by \begin{equation}
\label{btz-th} M(r_+)=\frac{r_+^2}{\ell^2},~C(r_+)=4\pi
r_+,~S_{BH}(r_+)=
 4\pi r_+,~F(r_+)= M-T_{\rm H} S_{BH}=-\frac{r_+^2}{\ell^2}.
 \end{equation}
We note that these quantities are positive regardless of the horizon size $r_+$ except
that the free energy is always negative.  This means that the BTZ
black hole is thermodynamically stable in Einstein gravity.  Also, it is easy to check
 that the first-law of thermodynamics is satisfied as
\begin{equation}\label{first-law}dM_{\rm ADT}=T_{\rm H} dS_{\rm ADT} \end{equation}
 as the
first-law is satisfied in Einstein gravity
 \begin{equation}
dM=T_{\rm H} dS_{\rm BH} \end{equation} where `$d$' denotes the differentiation
with respect to the horizon size $r_+$ only.  It is interesting to
observe that in the limit of $m^2 \to \infty$ one recovers
thermodynamics of the BTZ black hole in Einstein gravity, while in
the limit of $m^2 \to 0$ we recover the black hole thermodynamics in
purely fourth-order gravity.

We are in a position to discuss the linear stability of BTZ black hole in NMGdbs by considering metric perturbation $h_{\mu\nu}$ and scalar perturbation $\phi$ in  $g_{\mu\nu}=\bar{g}_{\mu\nu}+h_{\mu\nu}$ and $\Phi=0+\phi$.
Firstly, the linearized Einstein equation to (\ref{eqn}) upon
choosing the transverse-traceless  gauge ($\bar{\nabla}^\mu
h_{\mu\nu}=0$ and $h^\mu~_\mu=0$) leads to the fourth-order equation
for the metric perturbation
$h_{\mu\nu}$
\begin{equation}
 \Big[\bar{\Delta}_L+\frac{2}{\ell^2}
-{\cal M}^2(m^2)\Big]\Big(\bar{\Delta}_L+\frac{2}{\ell^2}\Big) h_{\mu\nu} =0,
\label{linh}
\end{equation}
where $\bar{\Delta}_L$ denotes the Lichnerowicz operator defined as
\begin{equation}
\bar{\Delta}_L h_{\mu\nu}=-\bar{\square}h_{\mu\nu}-\bar{R}_{\rho\mu\sigma\nu}h^{\rho\sigma}-\frac{2}{\ell^2}h_{\mu\nu}.
\end{equation}
This  might imply  the two second-order linearized equations
\begin{eqnarray}\label{nmgmeq1}
&&\Big(\bar{\Delta}_L+\frac{2}{\ell^2}\Big)h_{\mu\nu}=0,
\\ \label{nmgmeq2}
&&\Big[\bar{\Delta}_L+\frac{2}{\ell^2} -{\cal M}^2(m^2)\Big]h_{\mu\nu}^{\cal M}=0.
\end{eqnarray}
Here, the mass squared ${\cal M}^2$ of a massive spin-2 is given by Eq.(\ref{spin2-m}).
Importantly, Eq. (\ref{nmgmeq1}) describes a massless spin-2 (gauge degrees of freedom), whereas Eq. (\ref{nmgmeq2})
describes a massive spin-2  with 2 DOF propagating around the BTZ
black hole under the  transverse-traceless gauge.  Solving a coupled first-order equations for $s(n=0)$-mode $H_{tr}$  and $H_-=H_{tt}/f(r)-f(r)H_{rr}$ derived from Eq.(\ref{nmgmeq2}) with $h_{\mu\nu}=e^{\Omega t} e^{i n \varphi}H_{\mu\nu}$, the classical instability (Gregory-Laflamme instability) bound of the BTZ black hole on $m^2$ was found  by
\begin{equation}
{\cal M}^2<0 \to  m^2< \frac{1}{2\ell^2} \to m<m_{th},\quad m_{th}=\frac{0.7}{\ell} \label{GL-b}
\end{equation}
regardless of the horizon radius
$r_+$~\cite{Moon:2013lea}.

It is well known that  the local thermodynamic stability is
determined by the positive heat capacity ($C_{\rm ADT}>0$)  in NMGdbs.
For ${\cal M}^2>0(m^2>m^2_{th})$, all thermodynamic
quantities have the same property as those for Einstein gravity, whereas for ${\cal M}^2<0(m^2<m^2_{th})$, all
thermodynamic quantities have the same property as those for
fourth-order term.   We observe that for
${\cal M}^2>0$, the BTZ black hole is thermodynamically stable,
regardless of the horizon radius $r_+$ because of  $C_{\rm ADT}>0$.
The case of ${\cal M}^2=0(m^2=m^2_{th})$ corresponds
to the critical gravity where all thermodynamic quantities are
zero and logarithmic modes appear.
For ${\cal M}^2<0$, the BTZ black
hole is thermodynamically unstable because of $C_{\rm ADT}<0$  as
well as it is classically unstable against  metric perturbation.
 Hence, it shows a clear connection (CSC: correlated stability conjecture) between thermodynamic and
classical instability for the BTZ black hole regardless of the
horizon radius $r_+$ in NMGdbs. However, it is important to note that there is no such connection in Einstein gravity.

Finally, let us study the tachyonic instability for scalar perturbation.
By linearizing  Eq.(\ref{sca-eq}), one finds a linearized scalar equation with $\bar{R}=-6/\ell^2$
\begin{equation}
\bar{\square}\phi-\Big(-\frac{6\alpha}{\ell^2}+m^2_{\Phi}\Big) \phi=0. \label{lsca-eq}
\end{equation}
Considering the separation of variables as
\begin{equation}
\phi(t,\varphi,r)\propto e^{-i\omega t+ in \varphi} \frac{\psi(r)}{\sqrt{r}},
\end{equation}
one finds the Schr\"{o}dinger equation  with a tortoise coordinate defined by $ dr_*=dr/f(r)$ as
\begin{equation}
\frac{d^2\psi}{dr_*^2}+\Big[\omega^2-V_{\psi,\alpha}(r)\Big] \psi=0,
\end{equation}
where the scalar potential is given by
\begin{equation}
V_{\psi,\alpha}(r)=f(r)\Big[\frac{M}{4r^2}+\frac{n^2}{r^2}-\frac{6}{\ell^2}\Big(\alpha-\frac{1}{8}-\frac{m^2_\Phi \ell^2}{6}\Big)\Big]. \label{psi-p}
\end{equation}
The appearance of negative region in the far region  indicates tachyonic instability because the last term of Eq.(\ref{psi-p}) plays the role of an effective  mass term.

Observing Fig. 1, one finds that the tachyonic instability bound for $s(n=0)$-mode scalar is determined by
\begin{equation}
\alpha>\alpha_{th},\quad \alpha_{th}=\frac{1}{8}+\frac{m^2_\Phi \ell^2}{6}. \label{t-b}
\end{equation}
Consequently, we find two instability bounds (\ref{GL-b}) and (\ref{t-b}) from stability analysis of BTZ black holes in NMGdbs.
\begin{figure*}[t!]
   \centering
  \centering
  \includegraphics{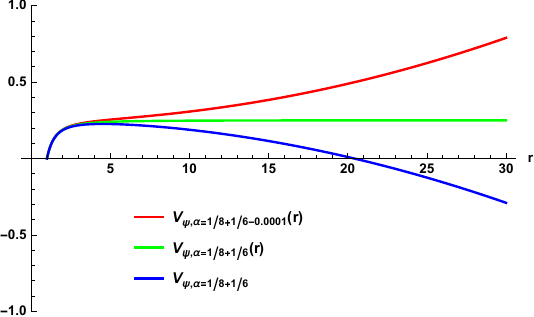}
\caption{ Scalar potential $V_{\psi,\alpha}(r\in[r_+=1,30])$ for $s(n=0)$-mode $\phi$  with $M=\ell^2=m^2_\Phi=1$ and $\alpha=1/8+1/6-0.0001$ (stable), $\alpha=1/8+1/6(=\alpha_{th})$ (threshold of tachyonic instability), and $\alpha=1/8+1/6+0.0001$ (unstable). The negative region increases as $\alpha$ increases because the last term of Eq.(\ref{psi-p}) plays the role of an effective  mass term. }
\end{figure*}
\section{Scalarized  black holes }
\begin{figure*}[t!]
   \centering
  \centering
  \includegraphics{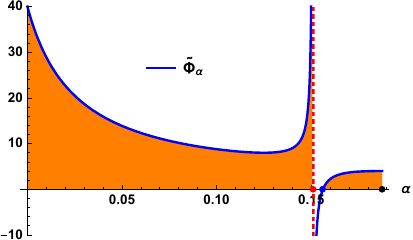}
  \hfill%
  \includegraphics{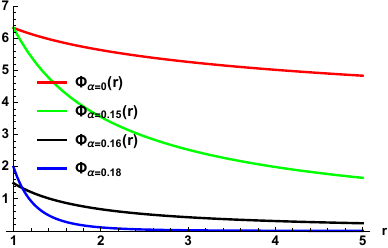}
\caption{(Left) $\tilde{\Phi}_\alpha$ as a function of $\alpha$ with $M=1$. The shaded region represents $\tilde{\Phi}_\alpha>0$, it blows up at $\alpha=1.151$ (dotted line) and it is zero at $\alpha=0.156,~0.1875$. (Right) Scalar hairs $\Phi_\alpha(r)$ as function of $r\in[r_+=1,5]$  with  $\ell=1$ and $\alpha=0,~0.15,~0.16,~0.18$.
The forbidden region is between $\alpha=0.151$ and $\alpha=0.156$. }
\end{figure*}

One  finds an analytic solution for scalarized black holes by solving (\ref{eqn}) and (\ref{sca-eq}) as~\cite{Correa:2014ika}
\begin{eqnarray}
ds_{\rm sbh}^2&=&-F(r)dt^2+\frac{dr^2}{F(r)}+r^2d\varphi^2,\quad F(r)=-M\Big(\frac{r}{\ell}\Big)^{\frac{32\alpha-5}{16\alpha-3}}+\frac{r^2}{\ell^2} \nonumber \\
\Phi_\alpha(r)&=&\Big(\frac{r}{\ell}\Big)^{\frac{0.5}{16\alpha-3}}\tilde{\Phi}_\alpha,\quad\tilde{ \Phi}_\alpha= \sqrt{\frac{8M(32\alpha-5)}{256\alpha^2-32\alpha-1}}, \label{sbh-s}
\end{eqnarray}
where $M$ denotes the ADM mass of scalarized black hole.
 We regard Eq.(\ref{sbh-s}) as non-BTZ black holes with scalar hair in three dimensions. One finds that $256\alpha^2-32\alpha-1= 256(\alpha-\frac{1+\sqrt{2}}{16})(\alpha-\frac{1-\sqrt{2}}{16})$.
Importantly, we observe that a case of $\alpha=\frac{5}{32}$ recovers the BTZ black hole  Eq.(\ref{btz-m}) with $\Phi=0$.
Imposing  a vanishing $\Phi_\alpha(r)$ at infinity and  a positive finite scalar hair on the horizon, we find two bounds for $\alpha$ (see Fig. 2)
\begin{equation}
0<\alpha<\frac{1+\sqrt{2}}{16}(=0.151),\quad \frac{5}{32}(=0.156)<\alpha<\frac{3}{16}(=0.1875). \label{phi-bo}
\end{equation}
For this solution, other parameters take the forms as
\begin{eqnarray}
m^2(\alpha)&=&\frac{256\alpha^2-32\alpha-1}{2(16\alpha-3)^2\ell^2}, \nonumber \\
\Lambda(\alpha)&=&-\frac{(16\alpha-1)(48\alpha-7)}{2(256\alpha^2-32\alpha-1)\ell^2}, \nonumber \\
m^2_\Phi(\alpha)&=& \frac{(8\alpha-1)(768\alpha^2-192\alpha+11)}{4(16\alpha-3)^2\ell^2}, \nonumber \\
\lambda(\alpha)&=& -\frac{3(8\alpha-1)(256\alpha^2-32\alpha-1)(768\alpha^2-152\alpha+9)}{16(16\alpha-3)^2(32\alpha-5)\ell^2}. \label{other-p}
\end{eqnarray}
Taking a positive $m^2(\alpha)$, one has a bound of $\alpha>\frac{1+\sqrt{2}}{16}$ [see (Left) Fig. 3].
For a negative  $\Lambda(\alpha)$, one has two bounds: $\frac{1}{16}(=0.063)<\alpha<\frac{7}{48}(=0.146)$ and $ \alpha>\frac{1+\sqrt{2}}{16}$[see (Right) Fig. 3]. Two roots of $768\alpha^2-192\alpha+11=0$ are given by $\alpha=0.089,0.161$. For a positive $m^2_\Phi(\alpha)$, we find two bounds:
$0.089<\alpha<\frac{1}{8}(=0.125)$ and $\alpha>0.161$ (see Fig. 4). However, there is no restriction on $\lambda(\alpha)$. A common bound from $m^2(\alpha)>0,~\Lambda(\alpha)<0,$ and $m^2_\Phi(\alpha)>0$ leads to
\begin{equation}
\alpha >0.161. \label{b-1}
\end{equation}

\begin{figure*}[t!]
   \centering
  \centering
  \includegraphics{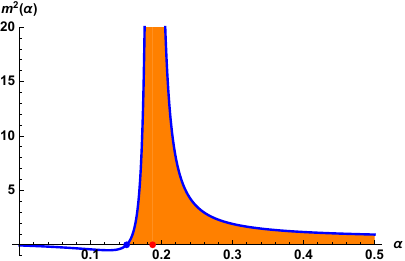}
  \hfill%
  \includegraphics{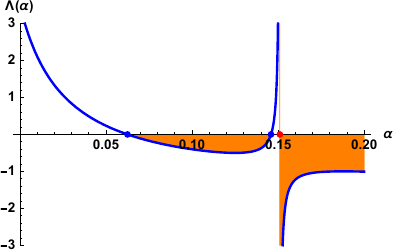}
\caption{(Left) Mass  $m^2(\alpha)$ as a function of $\alpha$ with $\ell=1$. The shaded region represents $m^2(\alpha)>0$ and it blows up at $\alpha=0.1875$ (dotted line). (Right) Cosmological constant $\Lambda(\alpha)$ as function of $\alpha$  with  $\ell=1$. The shaded region represents $\Lambda(\alpha)<0$ and it blows up at $\alpha=0.151$ (dotted line). }
\end{figure*}
\begin{figure*}[t!]
   \centering
  \centering
  \includegraphics{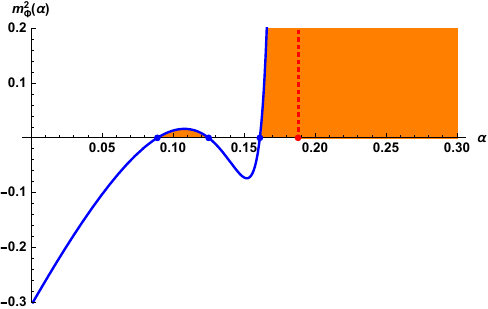}
\caption{(Left) Mass of scalar $m^2_\Phi(\alpha)$ as a function of $\alpha$ with $\ell=1$. Two shaded regions denote $m^2_\phi(\alpha)>0$ and it blows up at $\alpha=0.1875$ (dotted line).}
\end{figure*}
Finally, combining  Eq.(\ref{b-1}) with Eq.(\ref{phi-bo}), one finds a solution bound for obtaining scalarized black hole solutions
\begin{equation}
0.161 <\alpha^{\rm sol} <\frac{3}{16}(=0.1875).\label{b-2}
\end{equation}
Here, we obtain three isolated points for positive $\alpha$:  $\alpha_{\rm iso}=\frac{1+\sqrt{2}}{16},~ \frac{5}{32}, ~\frac{3}{16}$ because of $\alpha=\frac{1-\sqrt{2}}{16}<0$.
These all are excluded from the allowed parameter space.

\section{Complete analysis for thermodynamics of scalarized black holes}
First of all, the ($\alpha$-dependent) Hawking temperature is defined by
 \begin{equation}
  T_{\rm H}(r_+,\alpha)=
\frac{F'(r_+)}{4\pi}=\frac{r_+}{4\pi \ell^2(3-16\alpha)}.
 \end{equation}
Other ($\alpha$-dependent) thermodynamic quantities are found to be
 \begin{eqnarray}
\label{sbtz} M(r_+,\alpha)&=&\frac{r_+^2}{\ell^2}\Big(\frac{r_+}{\ell}\Big)^{\frac{32\alpha-5}{16\alpha-3}},\quad C(r_+,\alpha)=4\pi
r_+\Big(\frac{r_+}{\ell}\Big)^{\frac{32\alpha-5}{16\alpha-3}}, \nonumber \\
S_{BH}(r_+,\alpha)&=& 2\pi r_+\Big(\frac{16\alpha-3}{8\alpha-1}\Big)\Big(\frac{r_+}{\ell}\Big)^{\frac{32\alpha-5}{16\alpha-3}},\nonumber \\
 F_\alpha(r_+)&=& M-T_{\rm H} S_{BH}=\frac{r_+^2}{\ell^2}\Big(\frac{80\alpha-13}{16\alpha-2}\Big)\Big(\frac{r_+}{\ell}\Big)^{\frac{32\alpha-5}{16\alpha-3}}. \label{sbtz-th}
 \end{eqnarray}
 In deriving the entropy $S_{BH}$, we use the first-law of thermodynamics
 as
 \begin{equation}
 dM=T_{\rm H} dS_{BH}.
 \end{equation}
 For an isolated point ($\alpha=\frac{5}{32}$), we observe that the above reduces to  those in Eq.(\ref{btz-th}) for BTZ black hole exactly.
 For an isolated point ($\alpha=\frac{3}{16}$), any thermodynamic quantities are not defined properly because they blow up.
 Requiring that $T_{\rm H}(r_+,\alpha)>0 \to 0<\alpha<\frac{3}{16},~S_{BH}(r_+,\alpha)>0\to \frac{1}{8}<\alpha<\frac{3}{16},$ and $F_\alpha(r_+)<0\to \frac{1}{8}<\alpha<\frac{13}{80}$, one finds a thermodynamic  bound for obtaining  appropriate thermodynamic quantities for scalarized black hole as
 \begin{equation}
 \frac{1}{8}(=0.125)<\alpha^{\rm ther}<\frac{13}{80}(=0.1625). \label{b-3}
 \end{equation}
We note that in view of having  appropriately thermodynamic quantities,  two cases of  $\alpha=\frac{3}{16}$ and $\frac{1}{8}$ (stealth black hole) are  isolated points which are excluded from the allowed parameter space $\alpha$.
\begin{figure*}[t!]
   \centering
  \centering
  \includegraphics{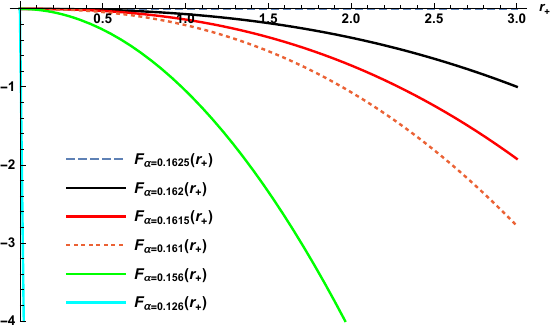}
\caption{ Helmholtz free energy $F_{\alpha}(r_+\in[0,10])$ with $\ell^2=1$ and $\alpha=0.1625$ (zero: $r_+$-axis), $\alpha=0.162$ (within bound (\ref{b-4})),  $\alpha=0.1615$ (within bound (\ref{b-4})), $\alpha=0.161$, $\alpha=0.156$ (BTZ), and  $\alpha=0.126$ ($F_\alpha$-axis). }
\end{figure*}

Consequently, combining Eq.(\ref{b-2}) with (\ref{b-3}), one has  a very specific bound for obtaining  scalarized black hole with appropriate thermodynamic quantities as
 \begin{equation}
0.161<\alpha<0.1625, \label{b-4}
 \end{equation}
  which is regarded as our key result.
We note that this bound belongs to the tachyonic bound (\ref{t-b}).
However, if the GL instability bound (\ref{GL-b}) is imposed, the mass bound on $m^2$ is modified as $\frac{1+\sqrt{2}}{16}<\alpha <\frac{5}{32}$.
In this case, we have  no such  bound like as Eq.(\ref{b-4}). In this sense, the GL instability bound might have noting to do with  constructing scalarized black holes.

Finally, we observe the Helmholtz free energy depicted in Fig. 5 to study phase transition between BTZ and scalarized black holes since all specific heats $C(r_+,\alpha)$ are positive definite.
We find that phase transition from BTZ black hole to scalarized black holes within the bound Eq.(\ref{b-4}) is unlikely to occur.
The BTZ black hole is always favored than scalarized black hole  because of $F_{\alpha=0.156(BTZ)}<F_{0.161<\alpha<0.625}$.
It needs to perform stability analysis of scalarized black holes within the bound Eq.(\ref{b-4}).

\section{Discussions}

We have investigated  the new massive gravity dressed by a nonminimally coupled scalar (NMGdbs) in Eq.(\ref{NMGAct}).
In this case, a quadratic scalar coupling to Ricci scalar $R$ was included, instead of a quadratic scalar coupling to fourth-order term $K=R_{\rho\sigma}R^{\rho\sigma}-3R^2/8$:
$S_{\rm NMGS}=\frac{1}{2} \int d^3x \sqrt{-g}~
[R-2\Lambda-(\partial\Phi)^2-(1-\Phi^2)K/m^2$ with $m^2$ a single massive spin-2 parameter. We note that  $K$ plays a role of the Gauss-Bonnet term in four dimensions because the Gauss-Bonnet term is identically  zero in three dimensions.
Even though the former coupling does not belong to the general setup like the latter coupling, we have considered the former coupling because it may admit an analytic black hole solution with scalar hair.

From the NMGdbs, we have found the non-rotating BTZ black hole as the bald black hole without scalar hair. We have made the linearized theory around the BTZ black hole to find the condition for instability and to trigger the spontaneous scalarization.
Linearizing two equations (\ref{eqn}) and (\ref{sca-eq}) leads to two linearized equations: one is the linearized Einstein equation and  the other is the linearized scalar equation. The linearized scalar equation (\ref{lsca-eq}) differs from the four-dimensional linearized scalar equation.
The mass term of the NMGdbs is a constant term like as `$-6\alpha/\ell^2+m^2_\Phi$', whereas the mass term of the EGBS theory is given by $\mu_{\rm S}^2=-\frac{96\alpha M^2}{r^6}$. The former induces negative potential in the far region, while the latter provides negative potential in the near horizon. Also, the linearized Einstein equation (\ref{nmgmeq2}) in  NMGdbs  is the same form as in NMG, whereas the linearized Einstein equation in the EGBS theory is the same as that for Einstein theory.
We have obtained the Gregory-Laflamme instability bound ($m<m_{th}=0.7/\ell$) and tachyonic instability bound ($\alpha>\alpha_{th}=1/8+m^2_\Phi \ell^2/6$) of bald BTZ black hole by making use of the linearized theory around BTZ black hole.

Solving two full equations  (\ref{eqn}) and (\ref{sca-eq}) directly, one has found  an analytic ($\alpha$-dependent) scalarized black hole (\ref{sbh-s}) with four parameters  $m^2(\alpha),~\Lambda(\alpha),~m^2_\Phi(\alpha),$ and $\lambda(\alpha)$.
We have obtained a solution bound of $0.161<\alpha^{\rm sol} <3/16$ for achieving  black holes with scalar hair analytically  and a thermodynamic bound of $1/8<\alpha^{\rm ther} <13/80$ for finding consistent thermodynamic quantities.
Without imposing the GL instability bound ($m<\frac{0.7}{\ell}$),  we have found a very narrow  bound of $0.161<\alpha<0.1625$ which is located inside tachyonic instability bound of $\alpha>\alpha_{th}$ with $\alpha_{th}=1/8+m^2_\Phi \ell^2/6$ for obtaining scalarized black holes.  However, if the GL instability bound is imposed, the mass bound on $m^2(\alpha)$ is modified as $\frac{1+\sqrt{2}}{16}<\alpha <\frac{5}{32}$.
In this case, there is  no such  bound like Eq.(\ref{b-4}). In this sense, it seems that the GL instability bound has noting to do with  constructing scalarized black holes.

On the other hand, the GL instability supports  correlated stability conjecture (CSC) between thermodynamic instability  and
classical   instability for bald  BTZ black hole regardless of the horizon radius $r_+$ in NMGdbs.

Consequently, the scalarized ($\alpha$-dependent) black hole exists within a very narrow band of $0.161<\alpha<0.1625$ in NMGdbs.

\end{document}